\documentclass[twocolumn,showpacs,showkeys,superscriptaddress]{revtex4}

\usepackage{amsmath}
\usepackage{bbold}
\usepackage{amsfonts}
\usepackage{amssymb}
\usepackage{pbsi}
\usepackage[T1]{fontenc}
\usepackage{hyperref}
\usepackage{xcolor}
\usepackage{graphicx}
\usepackage{subfig}
\usepackage{float}

\usepackage{color}

\newcommand{\comentario}[1]{{\bfseries\color{red} #1}}

\begin{document}

\title{Accelerating Airy tensor modes of cosmological gravitational waves}

\author{Claudio Aravena-Plaza}
\email{c_aravena@ug.uchile.cl (corresponding author)}
\affiliation{Departamento de F\'{\i}sica, Facultad de Ciencias, Universidad de Chile,
Santiago 7800003, Chile.}

\author{V\'ictor Mu\~noz}
\affiliation{Departamento de F\'{\i}sica, Facultad de Ciencias, Universidad de Chile,
Santiago 7800003, Chile.}

\author{Felipe A. Asenjo}
\affiliation{Facultad de Ingenier\'ia y Ciencias,
Universidad Adolfo Ib\'a\~nez, Santiago 7491169, Chile.}

\date{\today}

\begin{abstract}
From a classical analysis, it is shown that the nondiffractive
accelerating gravitational Airy wave packets are solutions of Einstein
equations for their linearized tensor modes in a
Friedmann-Lema\^itre-Robertson-Walker cosmological background
filled with a perfect fluid, with  equations of state  $w=1/3$ and
$w=-1/3$.  These solutions have finite energy, presenting accelerating
behavior due to the structured spatial form of the wavepacket. This is manifested by curved trajectories along the wave path.
Also,  using  spectral functions, it is possible, with these packets, to construct more general, arbitrary wave packets. All these new solutions bring insights on new forms for gravitational wave propagation. 
\end{abstract}


\maketitle

\section{Introduction}

Nondiffractive wave packets are known solutions of the  Schr\"odinger equation  \cite{berry, Yuce, Unnikrishnan} and Maxwell's equations \cite{kaminer, Hacyan}. Several experimental studies have been performed to show them as  physical modes, for instance, for light propagation \cite{kaminer, Hacyan, Aleahmad, Chremmos, Efremidis, Patsyk, Bandres, Esat}, fluids \cite{shenhe}, sound \cite{zhao2014delivering, chen2019broadband, tang2021generation}, heat diffusion \cite{functions2004applications, asenjo2021accelerating, asenjo2022airy} and relativistic electron plasmas \cite{winkler2023exact}. In gravitational waves, these kinds of solutions have also been obtained in flat spacetimes \cite{Asenjo}. The solutions in terms of Airy modes represent  accelerated wavepackets without diffraction in their trajectory, due to a non-vanishing Bohm potential \cite{hojman2020classical}, and thus, they describe solutions different to those described through plane waves.

On the other hand, from the linearization of the Einstein equations for a Friedmann-Lema\^itre-Robertson-Walker (FLRW) cosmological background metric, the dynamical equations for the scalar, vector and tensor modes of gravitational perturbations are straightforwardly obtained \cite{misner}. The tensor modes are of particular importance because they describe gravitational waves on this cosmology. In the simplest case, those  tensor modes propagate in a Universe filled with a perfect fluid with a specific equation of state. Depending on that, solutions for different stages of the Universe are modeled \cite{weinberg}. The detection of these waves is important due to the idea of an inflationary early Universe producing primordial gravitational waves \cite{osti_981884}.

In general, the study of tensor modes for an FLRW background metric can be performed exactly. Their dynamical evolution is determined by the equation of state of the cosmological content of the Universe, mainly described by a perfect fluid with pressure $p$ and its energy density $\rho$, and thus, an equation of state $w=p/\rho$. The different values of $w$ characterize the evolution of the scale factor of the Universe. The distinction is usually made between oscillatory solutions with a cutoff value of $w >-1/3$, where the solutions oscillate and damp, and with $w < -1/3$, where the oscillation occurs and then freezes at some value.

In this work, we focus in cosmologies with specific values $w=1/3$ and $w=-1/3$. In the following, we show that in these two particular cosmological scenarios, the propagation of gravitational waves can be described in terms of Airy functions. This behavior is due only to the structured spatial form of the wave, transverse to its propagation along  cosmological light-cone coordinates. This produces accelerating curved (parabolic) trajectories for the gravitational wave propagation.
Both values of $w$ here considered define different forms in which the Universe evolves.
The characterization of a Universe with $w=1/3$ represents a stage of the Universe where ultrarelativistic particles predominate. 
In this case, primordial gravitational waves can be found when Einstein linearized equations are coupled with Vlasov's equations  \cite{khlebnikov1984long}. On the other hand, a Universe with $w=-1/3$ falls within the so-called dark energy cosmologies, and consists of linear-type forms of energy, called the cosmic string. A cosmic string could arise as a topological defect that occurred during an early phase transition in the Universe. The NanoGrav Collaboration has recently reported strong evidence of common spectrum process, which was interpreted as a stochastic gravitational wave background (SGWB) in the framework of cosmic string \cite{PhysRevD.107.042003}. The importance of this measurement opens the field of observation of primordial background waves \cite{nanograv}. Furthermore, cosmic strings can produce primordial gravitational waves, such as gravitational wave bursts and SGWB \cite{Siemens_2007, https://doi.org/10.48550/arxiv.2207.03510, https://doi.org/10.48550/arxiv.2006.13872, CMB, Emond_2022,Caprini_2018}. In general, to show that a Universe with a cosmic string can produce primordial gravitational waves, different models described by some action are used, for example, using the Nambu-Goto action \cite{auclair2021window}  or Lagrangian models \cite{Emond_2022}.  Unlike those works, here we perform a tensor modal analysis finding Airy-type wave solutions.

The studied gravitational wave modes in this work correspond to spacetime perturbations of   the FLRW  background metric, which is described by the spacetime interval
\begin{eqnarray}
\label{eq:1}
ds^{2}=g_{\mu\nu}dx^{\mu}dx^{\nu}=a^2\left(-d\eta^2+\gamma_{ij}dx^{i}dx^{j}\right)\, ,
\end{eqnarray}
where $a=a(t)$ is the scale factor of the Universe, and 
$x^\mu=(t,x^i)$ (with $\mu,\nu=0,1,2,3$, and $i,j=1,2,3$). We have chosen the speed of light $c=1$. We can describe the dynamics in terms of the conformal cosmological time $\eta=\int dt/a$. The three-dimensional spatial metric $\gamma_{ij}$ can be represented in spherical coordinates as  
$\gamma_{ij}dx^{i}dx^{j}=dD^{2}+D_{A}^{2}d\Omega$,
where $D$ is the comoving distance, $D_{A}=K^{-1/2}\sin(K^{1/2}D)$ is the angular diameter distance, and $d\Omega$ is the differential angular element. 
$K$ is a constant representing the spatial curvature of the Universe ($K=0$ for spatially flat,   $K=1$ for closed Universe, and $K=-1$ for open Universe).

From the above metric, the gravitational tensor modes $H_{T}$ can be straightforwardly obtained \cite{weinberg}. They are 
obtained as a perturbation of the spatial components of metric \eqref{eq:1}.
By defining the effective gravitational wave tensor mode $Y$ (for any  component) by the relation 
$H_{T}=Y/a$, we get the following  equation for the cosmological gravitational tensor modes (see Appendix):
\begin{equation}
\label{eq:2}
-\nabla^{2}Y+ \ddot{Y}+\left(2K -\frac{\ddot{a}}{a}\right)Y=0\, ,
\end{equation}
where the overdot stands for the conformal time derivative, $d/d\eta$. 
Also, the tensor $Y$ must satisfy a gauge condition $\partial_{\beta}Y^{\beta}_{~\alpha}=0$.

Furthermore,  the background FLRW metric evolves following the equation ${\ddot{a}}/{a}=({4\pi G}/{3})a^{2}\rho(1-3 w)-K$, where $G$ is the gravitational constant. The fluid content satisfies the continuity equation $\dot{\rho}/\rho=-3\left(1+w\right)\left(\dot{a}/a\right)$, whose solution is 
$\rho =\rho_{0}a^{-3(1+w)}$, where $\rho_0$ is the initial energy density. 
Substituting all the above in Eq.~(\ref{eq:2}) for a spatially flat Universe with $K=0$, we finally obtain 
\begin{equation}
\label{eq:3}
-\nabla^{2}Y+\ddot{Y}-\frac{4\pi G}{3}a^{-1-3w}\rho_{0} (1-3w)Y=0\, .
\end{equation}

The solution of this equation can be found exactly in terms of Bessel functions for any equation of state
different to $-1/3$ and $1/3$.

\section{Airy  cosmological gravitational waves}

 The main purpose of this work is to show that, from Eq.~(\ref{eq:3}), Airy-type wavepacket solutions are obtained when $w=1/3$ and $w=-1/3$. This is performed in the following  sections, showing and discussing the different possible solutions. In both cases, the wave propagates along curved trajectories in the  plane formed by a cosmological conformal light-cone coordinate and a transverse direction.

\subsection{Accelerating  gravitational waves for $w=1/3$}

For  the case of the equation of state for ultra relativistic matter, wave equation \eqref{eq:3} reduces to  
\begin{equation}
 -\nabla^{2}Y + \ddot{Y}=0 \ .
 \label{eq:4}
\end{equation}

It is customary to solve this equation in terms of a sinusoidal basis.
By considering a polarized wave   given by
$Y=Y_{zz}=Y_{zz}(\eta,x,y)$, which satisfies the Lorenz gauge
condition $\partial_{z}Y_{zz}=0$, we can solve \eqref{eq:4} as a plane
wave $Y=\exp(i\omega \eta-ik_x x-i k_y y)$, such that
$\omega=\sqrt{k_x^2+k_y^2}$, implying a light-cone propagation of the
wave. This is achieved by the symmetric aspects of spatial coordinates
in this solution. However, other solutions can be constructed where
the different spatial coordinates are not  treated in the same
footing. These solutions lead to an accelerated propagation of the
wave, whose amplitude can be written in terms of Airy functions. For this, we assume that the polarized wave  has the form  
\begin{equation}
Y=Y_{zz}(\eta,x,y) = \Upsilon(\zeta,y)\exp(ik\tau),
\end{equation}
where $k$ is an arbitrary constant playing the role of a wavenumber. Also, we have defined the light-cone coordinates
\begin{eqnarray}
    \zeta=x-\eta\, ,\qquad
    \tau=x+\eta\, .
\end{eqnarray}
  Using this ansatz in Eq.~\eqref{eq:4}, the equation for $\Upsilon$ is
\begin{equation}
4ik\frac{\partial \Upsilon}{\partial \zeta} + \frac{\partial^{2} \Upsilon}{\partial^{2}y}=0\, ,
\label{eq:6}
\end{equation}
which is equivalent to a Schr\"odinger equation for a free particle.
It is very well-known that a possible solution of this equation can be
written in terms of Airy functions ${\mbox{Ai}}$ \cite{berry}, being
\begin{equation}
\Upsilon(\zeta,y)=\text{Ai}\left(2ky-k^{2}\zeta^{2}\right)\exp\left(2ik^{2}y-\frac{2i}{3}k^{3}\zeta^{3}\right)\, .
\label{eq:7} 
\end{equation}
This solution represents a nondiffractive accelerating gravitational wave packet, such that its intensity  bends its trajectory in a parabolic curve in the $\zeta$-$y$ plane. 

However, solution (\ref{eq:7}) has infinite energy. A finite energy Airy gravitational wavepacket has already been studied in Ref.~\citep{Asenjo}. Thus, for this case, the cosmological gravitational wave, with finite energy, 
is given by 
\begin{multline}
\Upsilon(\zeta,y)= \text{Ai}\left(2ky-k^{2}\zeta^{2}+ibk\zeta\right) \\
 \times \exp\left(bky-bk^{2}\zeta^{2} +\frac{i}{4}b^{2}k\zeta +
   2ik^{2}y\zeta-\frac{2i}{3}k^{3}\zeta^{3}\right),
\label{eq:8}
\end{multline}
in terms of an arbitrary real parameter $b$ that allows 
the normalization of the solution. This wavepacket has finite energy,
since $\int dy |\Upsilon|^{2}=\int dy~\Upsilon^{*}\Upsilon$ is independent of $\zeta$. Then, it is obtained that the wave packets are square integrable along $\zeta=0$ $(x=\eta)$ as
\begin{equation}
\int_{-\infty}^{\infty}dy |\Upsilon(\zeta,y)|^{2}=\int_{-\infty}^{\infty}dy |\Upsilon(0,y)|^{2}=\frac{e^{b^{3}/12}}{4k\sqrt{b\pi}},
\end{equation}
implying its finite energy. 

\begin{figure}
    \centering
    \includegraphics[scale=0.8]{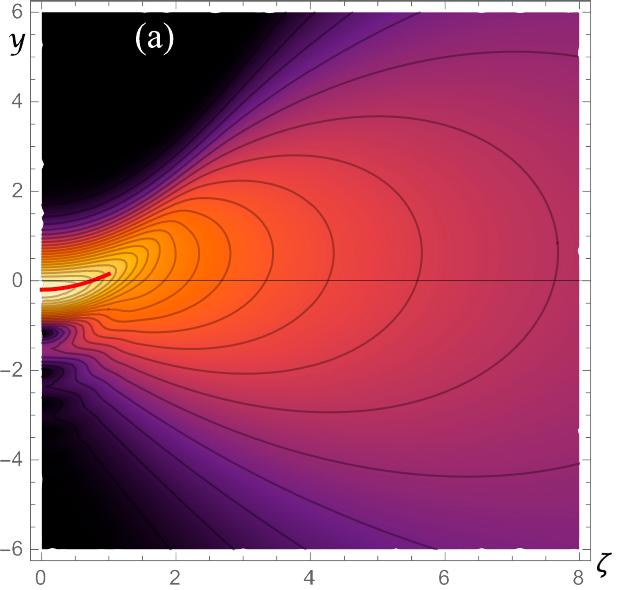}\\
    \includegraphics[scale=0.65]{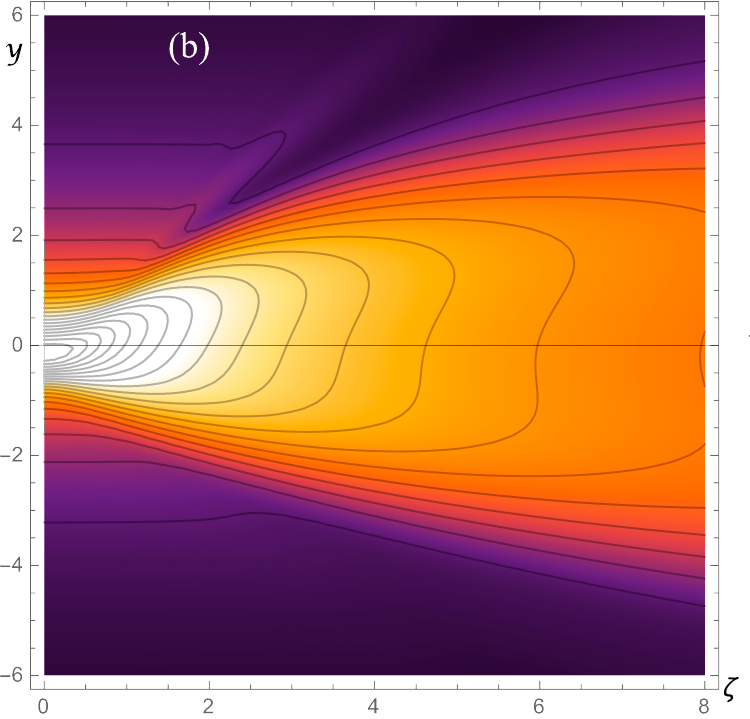}
    \caption{(a) Density plot for the magnitude  of $\Upsilon$, given by Eq.~\eqref{eq:8}, for $b=1$ and $k=1$. In red line, the parabolic trajectory of the maximum intensity of the solution is shown. (b) Density plot for the magnitude of generalized wavepacket ${\cal Y}(\zeta,y)$, given in Eq.~\eqref{genwavep},
 for spectral function $\varrho = \exp\left(-(k-k_0)^{2}\right)$, with $b=1$ and $k_0=1$.}
    \label{figura1}
\end{figure}

In Fig.~\ref{figura1}(a), we plot the magnitude of the solution \eqref{eq:8}, for $b=1$ and $k=1$. The parabolic trajectory followed by the intensity of the wave appears in the $\zeta$-$y$ plane. In red line,   the maximum maximorum of the solution \eqref{eq:8} is highlighted. For this case, and for small $\zeta$ values, the parabolic trajectory of this maximum is described by the approximate equation $y\approx {\zeta^2}/{3}+{\zeta}/{40}-{1}/{5}$, such that the initial acceleration of the wave, along this parabolic trajectory on the light-cone coordinates, is given by
\begin{equation}
    \frac{d^2y}{d\zeta^2}\approx \frac{2}{3}\, .
\end{equation}

The behavior of the different parts of the gravitational wavepacket \eqref{eq:8} is  non-local, making this solution very different to a plane wave. Instead of following straight lines in an isotropic space, the different parts of this solution (in particular, the maximum intensity of the gravitational wave)  bend its trajectory.

\subsection{Accelerating  gravitational waves for $w=-1/3$}

For this case of the equation of state, the Eq.~(\ref{eq:3}) becomes
\begin{equation}
 -\nabla^{2}Y + \ddot{Y} - \frac{8\pi G}{3}\rho_{0}Y=0.
 \label{eq:10}
\end{equation}
Thus, the content of the Universe acts as en effective imaginary mass for gravitational propagation. 

The Airy form for this propagation can be obtained using an ansatz
similar to the previous section. We consider a polarized wave $Y_{zz}$ with the form
\begin{equation}
Y=Y_{zz}(\eta,x,y)=\Upsilon(\zeta,y)\exp\left[ik\tau+ i S(\zeta)\right]\, ,
\end{equation} 
in terms of light-cone coordinates, and transverse coordinate $y$ (again, $k$ is an arbitrary constant). With this, Eq.~(\ref{eq:10}) reduces to 
\begin{equation}
4ik\frac{\partial \Upsilon}{\partial \zeta} + \frac{\partial^{2}\Upsilon}{\partial^{2}y}-4\left(k\frac{dS}{d\zeta} -\frac{2\pi G}{3}\rho_0\right)\Upsilon=0. 
\end{equation}
In order to have Airy-type wave packet solutions, the function $S$ must be chosen to be
\begin{equation}
 S(\zeta)= \frac{2\pi G \rho_0}{3k}\zeta\, ,
\end{equation}
to finally get 
\begin{equation}
4ik\frac{\partial \Upsilon}{\partial \zeta} + \frac{\partial^{2}\Upsilon}{\partial^{2}y}=0. 
\end{equation}
This equation is the same  as the one obtained for the $w=1/3$ case. Therefore, it has the same mathematical finite energy solution \eqref{eq:8}, evolving, for instance,  as it is depicted  in Fig.~\eqref{figura1}(a).

Therefore, the differences between the $w=-1/3$ and $w=1/3$ cases appear only on the phases of the gravitational waves, and not in their accelerating curved trajectories.

\section{Discussion on General wavepackets}

The Airy non-diffractive wavepacket solutions has been studied and measured in
quantum mechanics, electromagnetism and optics \cite{berry, Yuce, kaminer, Hacyan, Aleahmad, Efremidis, Patsyk, Bandres, Esat, Unnikrishnan}, and the solutions  presented in this work are the extension of this natural phenomena to cosmological gravitational waves. They  are particularly possible in the two distinctive Universes with $w=1/3$ and $w=-1/3$.

Importantly, the above solutions have a degree of freedom that can be
used to construct a more general set of wavepackets solutions of this
kind. This can be done by constructing the wavepacket as an average of
the previous solution through a spectral function that depends only on the  arbitrary parameter $k$. The general wavepacket acquires the form
\begin{equation}
    {\cal Y}(\zeta, y )= \int dk\, \varrho(k)\, Y(\zeta, y, k)\, ,
    \label{genwavep}
\end{equation}
where $Y(\zeta, y, k)$ is any of the solutions in the previous sections, and $\varrho(k)$ is an arbitrary spectral function, that can be chosen at will to define different wavepacket structures. Note that this wavepacket definition is clearly a solution of the equations (\ref{eq:4}) and \eqref{eq:10}, for the corresponding solutions for $Y$. Thus, this wavepacket correspond to an average on arbitrary distributions of different wavenumbers $k$.

As an appropriate example to demonstrate the dynamics of this general
wavepacket \eqref{genwavep}, we plot its magnitude  in Fig.~\eqref{figura1}(b), for
a spectral function determined by the Gaussian form
$\varrho(k)=\exp\left(-(k-k_0)^{2}\right)$.  This particular spectral
function selects only  wavenumbers close to $k_0$ as contributors to
the Airy dynamics. As a consequence, the Airy wavepacket becomes more
Gaussian as $\zeta$ progresses, such that the accelerating features of
the wave persist almost only for the maximum intensity lobe. This
example shows  how an almost-Gaussian accelerating gravitational wavepacket can be constructed from a set of Airy  wavepackets. In general, any desired form of accelerating gravitational wavepacket can be obtained by using an appropriate spectral function.

The two cosmological scenarios, for fluids with equations of state $w=1/3$ and $-1/3$, are not commonly considered into the study of cosmological gravitational waves. With this work, we hope to bring new insights on the accelerating features of the  gravitational wavepackets on those cosmological settings.

\begin{acknowledgments}
CAP thanks to ANID No.~7322/2020. This project has been financially supported by FONDECyT under
contracts No.~1201967 (VM), and  No.~1230094 (FAA). 
\end{acknowledgments}

\section{APPENDIX}

The gravitational tensor modes $H_{T}$ for a spacetime perturbation of the FLRW metric follow the wave equation \cite{weinberg}
\begin{equation}
\label{eq:19}
\nabla^{2}H_{T}-{a^{2}}\left(\dfrac{d^{2}}{dt^{2}}H_{T}\right)-3{a}\left(\dfrac{d}{dt}a\right)\left(\dfrac{d}{dt}H_{T}\right)=0\, .
\end{equation}
The evolution of these modes as a function of the cosmological time $\eta$ can be obtained by noticing that
$\dot{a}\equiv{da}/{d\eta}= a ({da}/{dt})$. Incorporating curvature constant $K$, we obtain (where $\dot{H}_{T}\equiv{d}H_{T}/d\eta$)
\begin{equation}
\label{eq:20}   
-\nabla^{2}H_{T} + \ddot{H}_{T} + 2\frac{\dot{a}}{a}\dot{H}_{T} + 2K H_{T}=0\, .
\end{equation}

\bibliographystyle{unsrt}
\bibliography{wave_airy}

\end{document}